# Aging Effects on Vertical Graphene Nanosheets and their Thermal Stability

Subrata Ghosh[1*], S. R. Polaki[1*], Aji Kumar[1], Nanda Gopala Krishna[2], and M. Kamruddin[1]

[1] *Surface and Nanoscience Division, Materials Science Group, Indira Gandhi Centre for Atomic Research-Homi Bhabha National Institute, Kalpakkam- 603102, India*
[2]*Corrosion Science and Technology Division, Metallurgy and Materials Group, Indira Gandhi Centre for Atomic Research, Kalpakkam-603102, India*

**Abstract**
Vertical Graphene Nano sheets (VGN) have become promising candidates for a wide range of applications due to their unique architecture and remarkable properties. However, utilization of VGN in potential applications relies on their structural and thermal stability in ambient conditions. The present study investigates the environmental aging effects and thermal stability of VGN. The effects of aging on VGN are investigated by Raman spectroscopy, X-ray Photoelectron spectroscopy and electrical measurements at regular intervals for a period of six months. Thermogravimetry was employed to examine the thermal stability of VGN and it is found that they are highly stable upto 575°C in ambient atmosphere. Detailed spectroscopic analysis substantiated the retention of structural quality and surface chemistry of VGN over the test period. These findings asserted the long-term structural as well as thermal stability of VGN up to 575°C and validate their potential utilization for the next-generation device applications.

*Corresponding authors: subrataghosh.phys@gmail.com (Subrata Ghosh), polaki@igcar.gov.in (S. R. Polaki)

## 1. Introduction

Vertical graphene nanosheets (VGN), is one of the novel architectures in graphene family, attracted a strong research momentum in a variety of applications spanning from energy storage to spintronics [1-4]. Remarkable properties of this unique nanoarchitecture that include high surface area, huge edge density, excellent electrical and thermal conductivity, easy functionalizability, bio-degradability etc., have drawn the attention of scientific community.[1, 5-7] Unlike graphene, a single layer hexagonal $sp^2$ bonded carbon parallel to the substrates, VGN are three dimensional open network of vertically oriented graphene sheets. It is also known as carbon nanowalls or carbon nanosheets. Each sheet of VGN contains a few layers of graphene with an interlayer spacing of around 0.36 nm. The thickness of the sheet is in nanometer scale, whereas its length and height varies up to several micrometers.

However, utilization of this material for device application relies on the stability against different environment and high temperatures. The VGN demonstrate excellent thermal stability with less than 3% weight loss upto 800 °C under $N_2$ environment.[8] Whereas high temperature annealing (1200-3000 °C) in high pure Argon (Ar) atmosphere found to modify the structure of VGN.[9] Our previous study had confirmed that the overall graphitization remains unchanged even after 16 hrs of annealing at 800 °C, but the edge modification and fragmentation occurs.[10] Wu *et al.*[11] reported the structural and morphological stability of VGN under harsh X-ray environment towards their potential application as cold cathode, even in the space. The reproducible field emission property and stability of VGN over 200 hrs of cycle operation at 1.3 mA emission current is demonstrated by Wang *et al.*[12]. PMMA/Graphene/VGN/Si solar cell shows a slight degradation after stored in air for 4 months by Jiao *et al.*.[13] Recent report showed retention of morphology and structural quality in transition metal oxide decorated VGN.[4] Many reports found the possibility to preserve VGN internal structure, architecture in different environments like acid solution, water and alcohol.[2, 14, 15] In addition to the stability against different environments, aging is one of the key and limiting factors for the utilization of materials in practical application. A Fourier Transform Infra-Red spectroscopic study of VGN, grown with higher Ar flow, shows a significant aging behavior, which is attributed to the deformation and stretching of the $sp^2$ and $sp^3$ vibration bands.[16] Detailed reports on investigation of long-term structural stability and electrical properties of VGN in ambient environment are missing. In view of the fact that aging can cause macro or microscopic structural and compositional changes, the study of long-term stability of VGN and its thermal stability in ambient conditions are of highly scientific interest.

The focus of the present study lays on investigating the effect of aging on the morphology, structure, chemical composition and electrical properties of VGN. A thorough investigation of these properties with respect to aging time is carried out at in periodic intervals over six months. Additionally, thermal stability of the VGN in ambient conditions was demonstrated by thermogravimetric analysis.

## 2. Experimental methods

*2.1 VGNs preparation*

The VGN grown on $SiO_2$ substrates by Electron Cyclotron Resonance Plasma Chemical Vapor Deposition (ECR-CVD) were used in this study. After loading the substrates, the ECR-CVD chamber was evacuated down to $10^{-6}$ mbar followed by substrate heating to 800 °C. Prior to the growth, substrates were annealed for 30 min and then cleaned by Argon plasma at 800 °C for 10 min. $CH_4$ (5N purity) at 5 sccm, as hydrocarbon source, was fed into the chamber along with the 20 sccm Ar (5N purity) as carrier gas. The flow rates of $CH_4$ and Ar were controlled by Alicat scientific mass flow controller. Growth was carried out at a Microwave power of 350 W, operating pressure of $10^{-3}$ mbar and temperature of 800 °C for 30 min. After the growth, microwave plasma and gas flow were turned off and allowed to anneal the as-grown VGN for 30 min at the same temperature. Finally, sample was cooled down to room temperature and taken out for further characterization.

*2.2 Characterization*

Field emission scanning electron microscopy (Supra 55, Zeiss, Germany) was employed to obtain the surface morphology of VGN. To assess the chemical and structural changes, Raman spectra of as-grown and aged VGN were recorded with 514 nm excitation in back scattering geometry with in-Via Renishaw Raman spectrometer, equipped with 50× objective lens (N.A. of 0.85) and 2400 gratings. Laser power was kept below 1 mW to avoid the laser induced heating. The chemical composition and surface chemistry of VGN were examined by X-Ray Photoelectron Spectroscopy (XPS) (SPECS Surface Nano Analysis GmbH) by employing Al Kα radiation (E = 1486.71 eV). The water contact angle of VGNs surface was measured by sessile drop method with the help of a CCD camera (Apex Instrument Co. Pvt. Ltd., India). The volume of the droplet is about 1μl and all measurements were carried out in ambient conditions. The value of contact angle is evaluated by half angle fitting method provided with the instrument.

Sheet resistance was measured using Agilent B2902A precision source/measure unit by four-probe resistivity method. Contacts were made by Ag paste and thin Cu wire.

Thermal stability was monitored using Thermo-gravimetric analyzer (SETSYS 16/18, SETARAM, France) equipped with high temperature furnace. Samples for TGA are prepared by peeling (scratching) out VGN from the substrate. Samples were taken in alumina crucibles and subjected to a heating rate of 10 °C/min from 25 to 800 °C under air atmosphere.

## 3. Results and Discussions

*3.1 Morphological analysis*

The scanning electron micrograph of VGN, shown in Fig. 1(a), and confirms its three-dimensional open network. The cross sectional micrographs indicate that the network is vertically standing on the substrate (inset of Fig. 1). The vertical height of VGN is measured to be 165 nm. Figure 1(b) confirms no variation in the VGN morphology and vertical height even after 180 days.

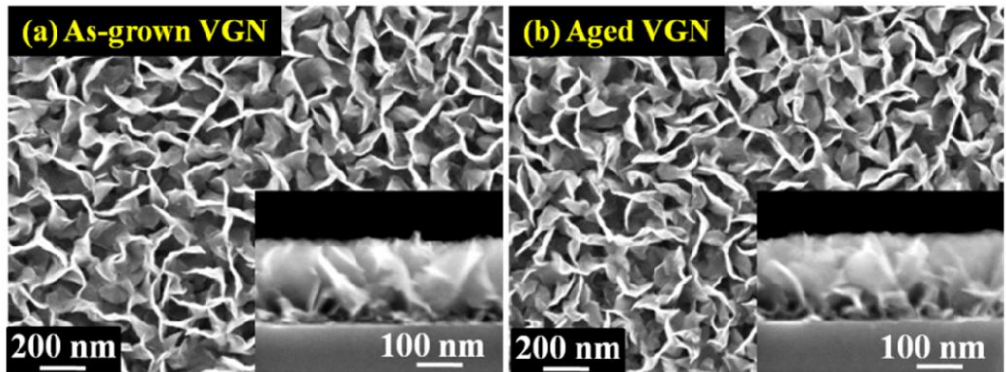

**Figure 1:** *Scanning electron micrographs of (a) as-grown and (b) 180 days aged vertical graphene nanosheets (VGNs). Insets represent the cross sectional micrograph of VGN.*

### 3.2 Raman spectroscopic study

Figure 2 shows typical Raman spectrum of VGN. The Raman spectrum of VGN has 8 peaks in the frequency range of 1000-3000 cm$^{-1}$, located at 1100, 1355, 1586, 1621, 2463, 2702, 2947 and 3238 cm$^{-1}$.[17] The presence of G-band at 1580 cm$^{-1}$ and G´-band at 2700 cm$^{-1}$ confirms the presence of graphitic structure. The 1355 cm$^{-1}$ peak is assigned as D-band. The G-band is originated due to the $E_{2g}$ stretching mode of $sp^2$ bonded carbon, whereas the D-band is due to the $A_{1g}$ breathing mode of $sp^2$ bonded carbon in rings.[18, 19]

The double resonance or triple resonance is responsible for the G´-band.[19] High intense D-band in the Raman spectra of VGN, as shown in Fig. 2(a), is attributed to the high edge density, ion-bombardment effect and vacancy-type defects.[20] The presence of D"-band (1100 cm$^{-1}$), D'-band (1621 cm$^{-1}$), D+D" (2463 cm$^{-1}$), D+D' (2947 cm$^{-1}$) and 2D' (3238 cm$^{-1}$) confirm the defective nature of VGN. The parameters of interest are (i) peak position of D-, G- and G'-bands, (ii) full width at half maximum (FWHM) of D-, G- and G'-bands and (iii) intensity ratio of D-to-G ($I_D/I_G$), D-to-D' ($I_D/I_{D'}$) and G'-to-G ($I_{G'}/I_G$).[18, 19] The peak position and FWHM of these three prominent peaks determines the structural quality, induced strain, effect of doping, disorder and number of graphene layers, whereas the intensity ratio of $I_D/I_G$, $I_D/I_{D'}$ and $I_{G'}/I_G$ are the measure of crystalline size, nature of defects and number of graphene layers, respectively.[18, 19]

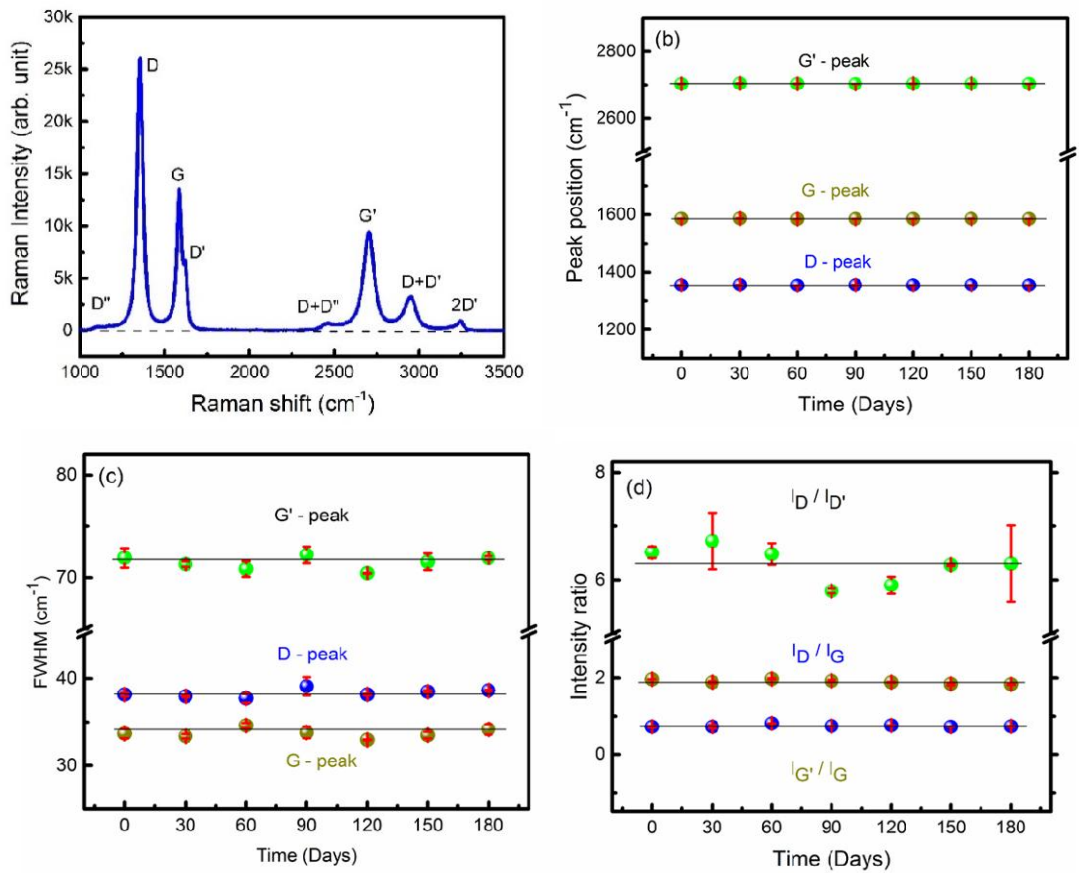

*Figure 2: (a) Typical Raman spectra of VGN. (b) peak position and (c) FWHM of D-, G- and G´-band of VGNs with respect to the aging time. (d) $I_D/I_G$, $I_D/I_{D'}$ and $I_{G'}/I_G$ versus aging time*

Raman spectra of the VGN are collected at three different positions in a regular interval of aging process for 180 days and the spectra are deconvoluted by Lorentzian lineshape. The average values of extracted parameters from the Raman spectra are plotted with error bars given in Fig. 2 denote standard deviation of certainty. The positions of D-, G- and G´-band, are found to be unchanged with respect to aging, as shown in Fig. 2(b). Moreover, Fig. 2(c) confirmed the unaltered FWHM of those prominent peaks against the aging time. To probe the sample further, the intensity ratios of D-to-G ($I_D/I_G$), D-to-D' ($I_D/I_{D'}$) and G'-to-G ($I_{G'}/I_G$) versus aging time are plotted in Fig. 1(d). The $I_D/I_{D'}$ ratio is found to vary within 5.8 to 6.5, it confirms the presence of vacancy-like defect in VGN. Each vertical sheet is composed of a few layers of graphene as it is reflected from the $I_{G'}/I_G$ ratio of 0.9. It can be clearly concluded here that the crystalline size, nature of defects and number of graphene layers are unaltered with time. Above experimental observations confirm the preservation of graphitic structure of VGN irrespective of time.

### 3.3 Surface chemistry

In order to evaluate the change in chemical composition of the surface upon aging, the surface chemistry of as-grown and aged VGN are investigated through XPS. The panels (a) and

(b) in Fig. 3 illustrate the C1*s* spectra of as-grown and aged VGN, respectively. The C1*s* spectra is deconvoluted by six Gaussian-Lorentzian line shape to identify the individual components of the carbon bonding and the best fit parameters are depicted as inset of Fig. 3(a) and (b). The major peaks at 284.00, 284.52 and 285.05 eV are assigned to vacancy-defect, C=C $sp^2$ and non-conjugated carbon, respectively.[21] The other less prominent peaks at 285.60, 286.30 and 287.05 eV are due to the presence of alcohol (C–OH), ether/epoxy, (C–O–C) and ketone/aldehyde (C=O) functional groups on the surface, respectively.[21] Since the edges of vertical sheets of VGN are partially H-terminated, they could form functional groups with atmospheric moisture/oxygen upon exposure to ambient conditions.[22] It is confirmed from the obtained deconvoluted result, inset of Fig. 3(a) and (b), that the binding energies of all peaks does not undergo any shift and there is no significant change in the carbon to oxygen ratio upon aging. This implies that the chemical nature of VGN and their exposed edges are quite stable despite aging time.

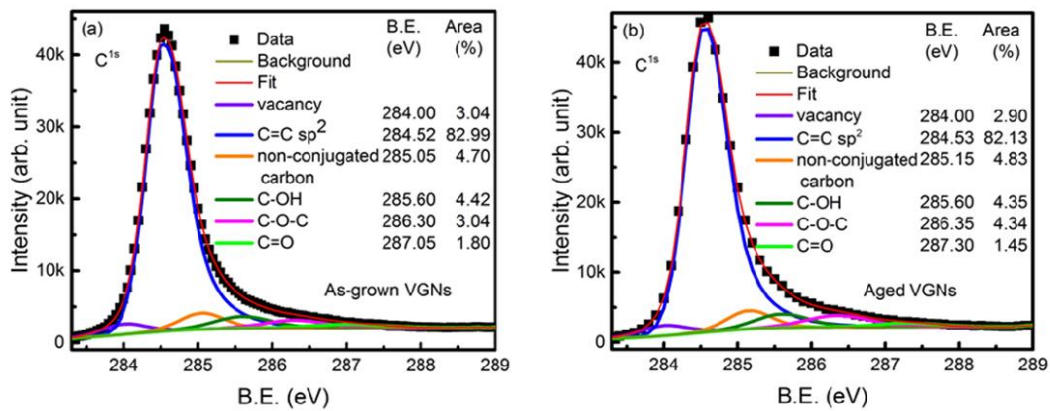

*Figure 3: C1s spectra of (a) as-grown and (b) aged VGN*

The water contact angle of as-grown VGN is found to be 133°, confirming their near super-hydrophobic nature without any modification of the surface (Fig. 4). The near superhydrophobic nature of VGN is due to its geometry and the wettability nature follows the Cassie model. Since the amount of oxygen functional group remains same for both the samples, the wetting nature does not undergo any significant change. Even after 180 days, the unchanged contact angle value shown in Fig. 4 confirms the preservation of morphology and surface chemistry. Such behavior is very important to utilize this material for self-cleaning, waterproof coating, anti-bio fouling and anti-corrosion properties, oil-water separator and micro-fluidic device.

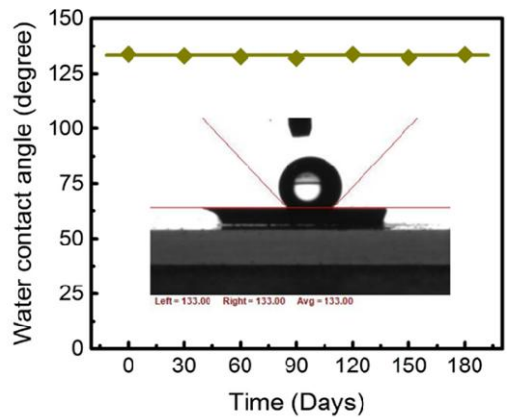

*Figure 4: Water contact angle of vertical graphene nanosheets*

*3.4 Electrical property*

In addition to the investigation on morphology, structure and chemical composition, it is essential to verify the influence of aging process on the electrical properties of VGN. With this motivation, the sheet resistance of VGN is measured at regular intervals for 180 days using four-point probe method. Considering VGN as a thin film in microscopic point of view, the sheet resistance is measured using van-der Pauw geometry.[5] The aging behavior in sheet resistance of VGN is shown in Fig.5. Inset of Fig.5 represent the current-voltage relationship of VGN, which confirms its Ohmic nature. The estimated sheet resistance of as-grown VGN is found to be 1.02 KΩ/□. As seen from the Fig.5, sample does not show any influence of aging on its sheet resistance value. The observed result affirms the excellent stability in electrical properties of VGN with respect to aging.

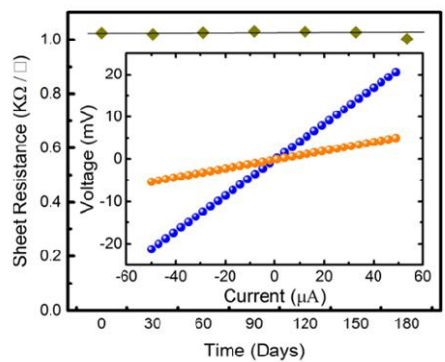

*Figure 5: Variation in sheet resistance of VGNs with aging time. Inset represents the I-V characteristics VGNs. Solid line is guide to eye*

*3.5 Thermal Stability*

Besides the long-term stability, thermal stability of VGNs is determined using TGA in ambient environment. TGA is a well-known technique to investigate the oxidative stability of carbon nanostructures.[23] Figure 6 depicts weight loss vs temperature of VGN, shows its thermal stability in ambient atmosphere. The initial 6% weight loss upto 250°C is attributed to the elimination of adsorbed water. Excellent thermal stability of VGN is exhibited up to 575 °C, as clearly seen from TGA profile (Fig. 6). The full weight loss of the sample above 575 °C is due to the complete oxidation of carbon in air ($C + O_2 = CO_2 \uparrow$). The high thermal stability of VGN in air is very significant for its utilization in potential applications.

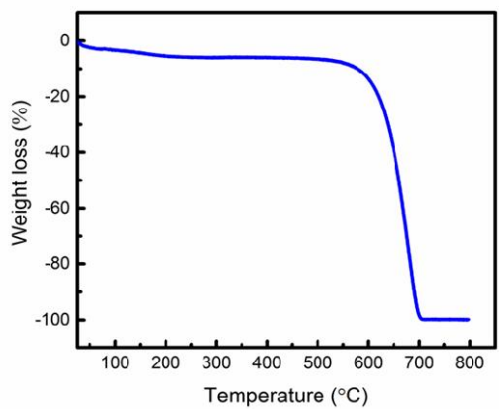

*Figure 6: Thermal stability of vertical graphene nanosheets in air.*

## 4. Conclusions

The morphological, structural, electrical property and thermal stability of vertical graphene nanosheets (VGN) are investigated using SEM, Raman spectroscopy, XPS, four-probe resistance measurement and thermogravimetric analysis, respectively. The novel architecture exhibited an excellent morphological, structural, and electrical stability over a period of 180 days. They also possess thermal stability in ambient conditions upto 575°C. The observed facts on long-term stability of VGN pave way for their potential utilization for device applications.

**Author information**

Author contributions
S.G. planned, performed the experiments, analyzed the data and wrote the manuscript. A.K. did the TGA experiment. N.G.K. performed XPS experiment. All authors discussed the results, commented on the manuscript and gave approval to the final version of the manuscript. S.G., S.R.P., A.K., N.G.K., and M.K. contributed to the revision of the manuscript.

**Notes**
The authors declare no competing financial interests.

**Acknowledgements**
We acknowledge Dr. S. Dhara for allowing to access Raman spectrometer and Dr. A. K. Tyagi for kind support. S.G. acknowledges Dept. of Atomic Energy, Govt. of India for his Postdoctoral Research Associate ship.

**References**

[1] J. Chen, Z. Bo, G. Lu, Vertically-Oriented Graphene, Springer, 2015.
[2] S. Ghosh, T. Mathews, B. Gupta, A. Das, N.G. Krishna, M. Kamruddin, Supercapacitive vertical graphene nanosheets in aqueous electrolytes, Nano-Structures & Nano-Objects, 10 (2017) 42-50.
[3] S.C. Ray, N. Soin, T. Makgato, C. Chuang, W. Pong, S.S. Roy, S.K. Ghosh, A.M. Strydom, J. McLaughlin, Graphene supported graphone/graphane bilayer nanostructure material for spintronics, Sci. Rep., 4 (2014) 3862.
[4] S. Ghosh, B. Gupta, K. Ganesan, A. Das, M. Kamruddin, S. Dash, A.K. Tyagi, MnO2-Vertical graphene nanosheets composite electrodes for energy storage devices, Mater Today Proc., 3 (2016) 1686-1692.
[5] H.J. Cho, H. Kondo, K. Ishikawa, M. Sekine, M. Hiramatsu, M. Hori, Density control of carbon nanowalls grown by CH4/H2 plasma and their electrical properties, Carbon, 68 (2014) 380-388.
[6] K.K. Mishra, S. Ghosh, R.R. Thoguluva, S. Amirthapandian, M. Kamruddin, Thermal Conductivity and Pressure Dependent Raman Studies of Vertical Graphene Nanosheets, J. Phys. Chem. C, 120 (2016) 25092–25100.
[7] N. Suetin, S. Evlashin, A.V. Egorov, K.V. Mironovich, S. Dagesyan, L.V. Yashina, E.A. Gudilin, V. Krivchenko, Self-assembled nanoparticle patterns on carbon nanowall surface, Phys. Chem. Chem. Phys., 18 (2016) 12344-12349.
[8] B. Shen, J. Ding, X. Yan, W. Feng, J. Li, Q. Xue, Influence of different buffer gases on synthesis of few-layered graphene by arc discharge method, Appl. Surf. Sci., 258 (2012) 4523-4531.
[9] Z. Wang, H. Ogata, S. Morimoto, M. Fujishige, K. Takeuchi, Y. Hashimoto, M. Endo, Structure changes of MPECVD-grown carbon nanosheets under high-temperature treatment, Carbon, 68 (2014) 360-368.
[10] S. Ghosh, K. Ganesan, S.R. Polaki, A.K. Sivadasan, M. Kamruddin, A.K. Tyagi, Effect of Annealing on the Structural Properties of Vertical Graphene Nanosheets, Adv. Sci. Eng. Med., 8 (2016) 146-149.
[11] J. Wu, Y. Zhang, B. Wang, F. Yi, S. Deng, N. Xu, J. Chen, Effects of X-ray irradiation on the structure and field electron emission properties of vertically aligned few-layer graphene, Nuclear Instruments and Methods in Physical Research Section B, 304 (2013) 49-56.
[12] S. Wang, J. Wang, P. Miraldo, M. Zhu, R. Outlaw, K. Hou, X. Zhao, B.C. Holloway, D. Manos, T. Tyler, High field emission reproducibility and stability of carbon nanosheets and nanosheet-based backgated triode emission devices, Appl. Phys. Lett., 89 (2006) 183103.
[13] T. Jiao, D. Wei, X. Song, T. Sun, J. Yang, L. Yu, Y. Feng, W. Sun, W. Wei, H. Shi, C. Hu, C. Du, High-efficiency, stable and non-chemical-doped graphene-Si solar cells through interface engineering and PMMA antireflection, RSC Adv., 6 (2016) 10175-10179.
[14] S. Evlashin, S. Svyakhovskiy, N. Suetin, A. Pilevsky, T. Murzina, N. Novikova, A. Stepanov, A. Egorov, A. Rakhimov, Optical and IR absorption of multilayer carbon nanowalls, Carbon, 70 (2014) 111-118.
[15] S. Vizireanu, G. Dinescu, L.C. Nistor, M. Baibarac, G. Ruxanda, M. Stancu, D. Ciuparu, Stability of carbon nanowalls against chemical attack with acid solutions, Physica E: Low dimens. Syst. Nanostruct., 47 (2013) 59-65.

[16] V. Marascu, S. Vizireanu, S. Stoica, V. Barna, A. Lazeastoyanova, G. Dinescu, FTIR investigation of the ageing process of carbon nanowalls, Romanian Reports in Physics, 68 (2016) 1108–1114.

[17] S. Ghosh, K. Ganesan, S.R. Polaki, T. Ravindran, N.G. Krishna, M. Kamruddin, A. Tyagi, Evolution and defect analysis of vertical graphene nanosheets, J. Raman. Spectrosc., 45 (2014) 642-649.

[18] A.C. Ferrari, D.M. Basko, Raman spectroscopy as a versatile tool for studying the properties of graphene, Nat. Nanotechnol., 8 (2013) 235-246.

[19] L. Malard, M. Pimenta, G. Dresselhaus, M. Dresselhaus, Raman spectroscopy in graphene, Phys. Rep., 473 (2009) 51-87.

[20] S. Ghosh, K. Ganesan, S.R. Polaki, S. Ilango, S. Amirthapandian, S. Dhara, M. Kamruddin, A.K. Tyagi, Flipping growth orientation of nanographitic structures by plasma enhanced chemical vapor deposition, RSC Adv., 5 (2015) 91922-91931.

[21] G. Karuppiah, S. Ghosh, N.G. Krishna, S. Ilango, M. Kamruddin, A.K. Tyagi, A comparative study on defect estimation using XPS and Raman spectroscopy in few layer nanographitic structures, Phys. Chem. Chem. Phys., 18 (2016) 22160-22167.

[22] H. Watanabe, H. Kondo, M. Hiramatsu, M. Sekine, S. Kumar, K. Ostrikov, M. Hori, Surface Chemical Modification of Carbon Nanowalls for Wide-Range Control of Surface Wettability, Plasma Process. Polym., 10 (2013) 582-592.

[23] D. Bom, R. Andrews, D. Jacques, J. Anthony, B. Chen, M.S. Meier, J.P. Selegue, Thermogravimetric Analysis of the Oxidation of Multiwalled Carbon Nanotubes: Evidence for the Role of Defect Sites in Carbon Nanotube Chemistry, Nano Lett., 2 (2002) 615-619.